\documentclass[twocolumn]{aastex62}
\turnoffedit
\usepackage{array}
\usepackage{graphicx}
\usepackage{color}
\usepackage{amsmath}
\usepackage{booktabs}
\usepackage{tabularx}
\graphicspath{{figs/}}
\usepackage{placeins}
\usepackage[english]{babel}
\usepackage{multirow}
\usepackage[autostyle, english = american]{csquotes}
\MakeOuterQuote{"}
\newcolumntype{P}[1]{>{\centering\arraybackslash}p{#1}}
\newcolumntype{R}[1]{>{\raggedleft\arraybackslash}p{#1}}
\newcolumntype{Y}{>{\raggedleft\arraybackslash}p{.41in}}

\newcommand\baselineFPR{77}
\newcommand\baselineFPRknown{47}

\newcommand\baselineYield{2.0}
\newcommand\baselineYieldknown{5.3}
\newcommand\baselineRmean{2.3}
\newcommand\baselineRmeanknown{1.7}
\newcommand\baselineamean{1.2}
\newcommand\baselineameanknown{1.1}

 \shorttitle{Earth twin false positives} 
 \shortauthors{Guimond \& Cowan}

\begin{document}

\title{The direct imaging search for Earth 2.0: Quantifying biases and planetary false positives}
\AuthorCallLimit=2
\correspondingauthor{Claire Marie Guimond}
\email{claire.guimond2@mail.mcgill.ca}

\author{Claire Marie Guimond}
\affiliation{Department of Earth \& Planetary Sciences, McGill University, 3450 rue University, Montr\'eal, QC Canada, H3A 0E8}

\author{Nicolas B. Cowan}
\affiliation{Department of Earth \& Planetary Sciences, McGill University, 3450 rue University, Montr\'eal, QC Canada, H3A 0E8}
\affiliation{Department of Physics, McGill University, 3600 rue University, Montr\'eal, QC Canada H3A 2T8}

\revised{\today}
\submitjournal{\aj}
%\date{\today}                                          

\begin{abstract}

Direct imaging is likely the best way to characterize the atmospheres of Earth-sized exoplanets in the habitable zone of Sun-like stars. Previously, \citet{Sta14, Sta15, Sta16} estimated the Earth twin yield of future direct imaging missions, such as LUVOIR and HabEx. We extend this analysis to other types of planets, which will act as false positives for Earth twins. We define an Earth twin as any exoplanet within half an $e$-folding of 1 AU in semi-major axis and  1 $R_\Earth$ in planetary radius, orbiting a G-dwarf. Using Monte Carlo analyses, we quantify the biases and planetary false positive rates of Earth searches. That is, given a pale dot at the correct projected separation and brightness to be a candidate Earth, what are the odds that it is, in fact, an Earth twin? Our notional telescope has a diameter of 10 m, an inner working angle of 3$\lambda/D$, and an outer working angle of 10$\lambda/D$ (62 mas and 206 mas at 1.0 $\micron$). With no precursor knowledge and one visit per star, 77\% of detected candidate Earths are actually un-Earths; their mean radius is 2.3 $R_\Earth$, a sub-Neptune. The odds improve if we image every planet at its optimal orbital phase\edit2{, either by relying on precursor knowledge, or by performing multi-epoch direct imaging}. In such a targeted search, 47\% of detected Earth twin candidates are false positives, and they have a mean radius of 1.7 $R_\Earth$. The false positive rate is insensitive to stellar spectral type and the assumption of circular orbits.

%The majority of false positives will be "big and dark" planets with large radii and low apparent albedos. Indeed, the radius-albedo degeneracy is the ultimate challenge in reflected light direct imaging. The false positive rate is $\gtrsim$50\% unless all planets have the same albedo and we know that value \textit{a priori}. We might reduce the degeneracy via a mass-radius relation, if we know planetary mass from radial velocity or astrometry. 

\end{abstract}


\section{Introduction}

Planned direct imaging missions would measure the reflectance spectra \citep{DeM02} and photometric variability \citep{For01} of Earth-sized planets orbiting in the habitable zone of nearby Sun-like stars. Many studies have shown that direct imaging is also a viable way to discover these planets \citep{Ago07, Sta16, Sta15, Sta14}. Given enough time, a mission could discover hundreds to thousands of planets and characterize them all. In practice, there will only be enough time to characterize some of these worlds in detail. We would therefore like to distinguish between Earths and un-Earths as efficiently as possible. For although they are expected to revolutionize many aspects of planetary science, mission concepts such as LUVOIR and HabEx are being motivated based on their ability to characterize Earth twins.

\citet{Bro05} presented a "photometric and obscurational single-visit completeness" method to estimate the \replaced{ratio of detectable planets to undetectable planets for a given set of target stars}{chance, for a particular star, that a companion exoplanet is detectable during one visit given that the planet exists}. In their model, "photometric" refers to the condition that the planet/star contrast must exceed the inherent instrument floor in photon counting. "Obscurational" refers to how the planet and its star must be positioned in the sky plane, such that the planet is outside the inner obscuring disk of the coronagraph or starshade. \added{This inner working angle (IWA) is defined technically as the angle at which transmission decreases by 50\%. Coronagraphs may also have an outer working angle (OWA), beyond which starlight is no longer adequately suppressed. }\replaced{These}{Obscuration and low contrast} are the two dominant factors that could hinder a detection.\footnote{\added{Others include exo-zodiacal dust \citep{Rob12} and integration time.}} 

If one is equally interested in all planets, then an "average" mission completeness suffices, without looking at the demographics of the mission's yield. But what if one prefers a certain kind of planet? Then we would do best to consider how a mission may \replaced{not be so complete for some}{be biased towards inopportune} radii and semi-major axes. 

While \citet{Sta14, Sta15, Sta16} cared about semi-major axes between 0.7\textendash1.5~AU, they assigned a radius of 1~$R_\Earth$ to all planets in their completeness calculator. In reality, most planets \replaced{do not have $R\sim R_\Earth$}{are not the size of Earth}. 

\replaced{And the crux: any mission capable of finding Earth twins will have an easier time finding other sorts of planets. Hence, w}{W}e want to not only detect as many Earth twins as possible, but also know that a detected planet is an Earth twin. Many un-Earths will show up at the correct projected separation and brightness to be Earthlike. We would confuse these planets with true Earth twins, so we call them false positives.

\begin{figure}
\epsscale{1.2}
    \plotone{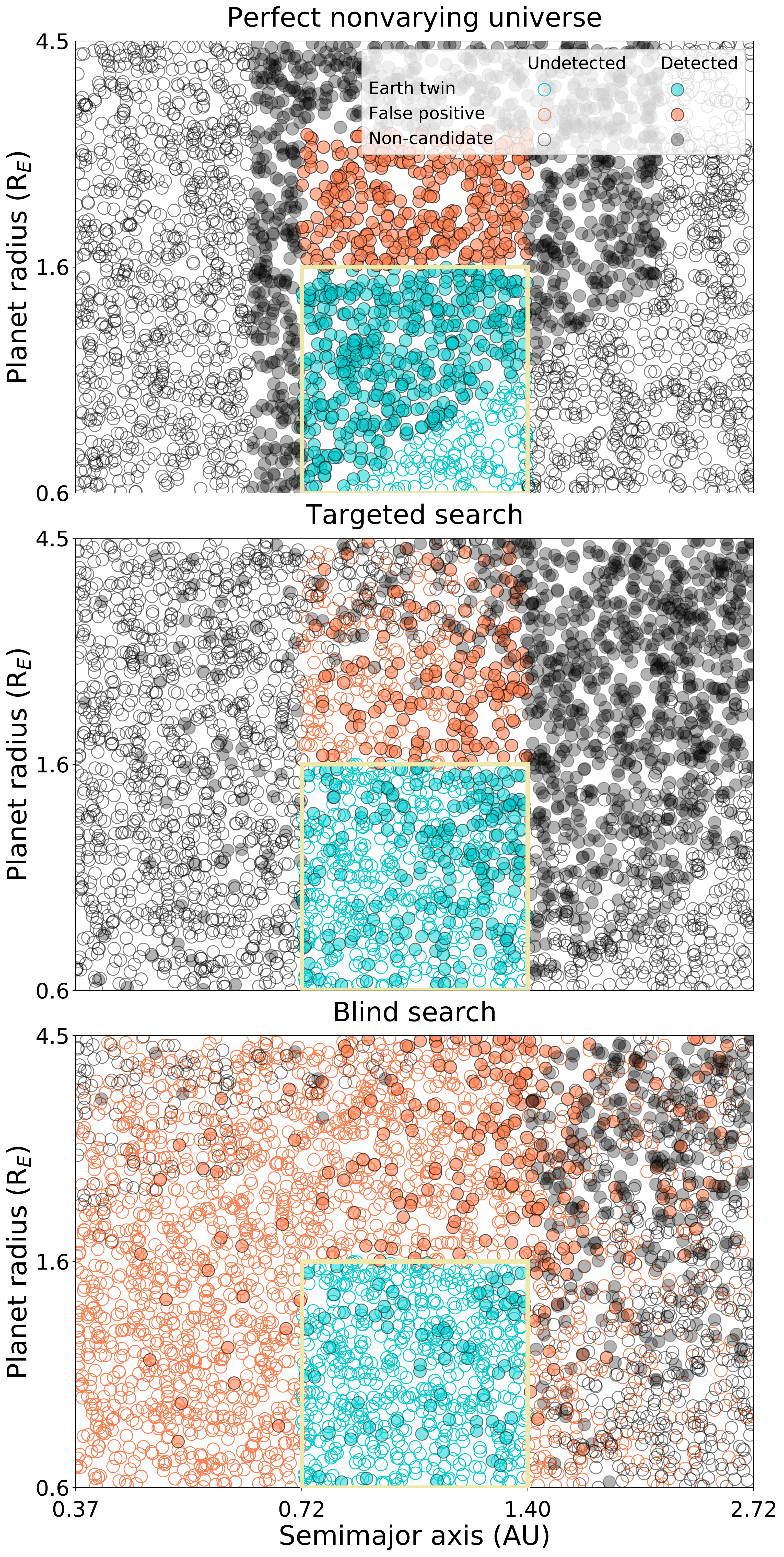}
    \caption{Demographics of detected (filled) and undetected (hollow) planets for three simulated surveys. Blue denotes an Earth twin, while orange denotes a false positive, and grey denotes a planet that would not be mistaken for an Earth twin. \textit{Top}: a survey of an ideal universe with every star at 10~pc, and planets with face-on inclination and 30\% albedo. \textit{Middle}: a search of a universe where \added{stellar }distance and orbital inclination vary randomly; albedo can uniformly vary from 0.05 to 0.5, and the planet is imaged at gibbous phase just outside the inner working angle. \textit{Bottom}: a search of a universe where distance, inclination, albedo, and orbital phase are random. The grid cell defining Earth twins is highlighted in yellow. Planets are distributed log-uniformly in semi-major axis and radius. Based on a simulation with stellar number density inflated by $\sim$1.5 orders of magnitude to $5\times10^3$ stars, for visualization.\label{fig:wedge}}
\end{figure}

\added{\subsection{An observation flowchart}

Suppose we image a star and see a dot that we have identified as a companion, and which may be a newly-discovered Earth twin. Our options include: (i) we get a spectrum of the dot immediately; or (ii) we return to this star at a later epoch, to better constrain the companion's semi-major axis, hoping that the planet has not become obscured by the IWA or confused with another planet in the system. If we choose option (ii), and the next image is not dissuading, then the choices are the same, \textit{ad infinitum} until we are ready to commit to spectroscopy.\footnote{\added{A silver lining to obtaining spectra of un-Earths is that they provide a control for biosignatures, as long as we eventually determine which planets are in fact habitable.}} A third option is to get another image in a different filter, if one believes that colour is a useful discriminant between different types of planets \citep{Kri16}, but phase-variable colours make this strategy more challenging \citep{Cah10,May16}. 

Roughly speaking, one direct imaging detection provides two data: the RA and Dec of the planet relative to its host star. There are seven orbital parameters, so $\gtrsim4$~detections are needed to establish an orbit. However, it is beyond our current scope to determine the best number of revisits, or their cadences. 

Rather, our analysis considers two endmember scenarios. In "blind" searches, we assume no prior observations of the planet; our only known parameters are the two first-order direct imaging observables of planet/star brightness contrast, $\varepsilon$, and projected separation, $a_{\rm proj}$. 

On the other hand, in "targeted" searches, we have the luxury of knowing where and when to look at each system. We assume their orbits can be predicted, based on data from either multi-epoch direct imaging, radial velocity,\footnote{\added{Radial velocity leaves two orbital parameters unconstrained: orbital inclination and longitude of the ascending node.}} or astrometry. The Keplerian orbital fits from these observations are adequate for us to target the wanderers at gibbous phase outside the IWA \citep{Sha18,But17,Ran17}.}

We therefore investigate how well a direct imaging mission can distinguish between Earths and un-Earths, based solely on photometry. Particularly, we focus on the \replaced{two lowest-order direct imaging observables: the planet/star contrast, and the projected separation}{"blind" and "targeted" observation scenarios}. In section 2, we describe our Monte Carlo method for simulating exoplanets and evaluating their detectabilities. Section 3 presents results, and section 4 our discussion\added{, including a sensitivity analysis to test our assumptions}. 

\section{Model description}

\subsection{Direct imaging signal scaling}

The signal from a directly imaged planet is the planet/star contrast ratio, parameterized for reflected light as
\begin{equation}\label{eq:CR}
\varepsilon = A^* \,\phi_L(\alpha) \left(\frac{R}{a}\right)^2, \\
\end{equation}
\noindent with planetary radius $R$, semi-major axis $a$, and apparent albedo $A^*$ \citep{Tra10}. The phase function $\phi_L(\alpha)$ describes how the \deleted{reflected }light scattered by a planetary atmosphere changes with \replaced{phase}{the star-planet-observer} angle $\alpha$. For the purposes of our numerical experiment, we adopt the Lambertian phase function:
\begin{equation}\label{eq:phi}
\phi_L(\alpha) = \frac{1}{\pi}\left[\sin{\alpha} + \left(\pi - \alpha \right)\cos{\alpha}\right].
\end{equation}
Th\replaced{is}{e} phase angle \deleted{between the host star, its planet, and the observer }is related trigonometrically to orbital phase $\xi$ and inclination $i$ \citep{Tra10}:
\begin{equation}\label{eq:alpha}
\alpha = \cos^{-1} \left(\cos{\xi} \sin{i}\right).
\end{equation}

\subsection{Conditions for detectability}

\replaced{Now for illustration---t}{Figure \ref{fig:wedge} illustrates the detection \replaced{biases in}{conditions for} direct imaging. T}he top panel \deleted{of figure \ref{fig:wedge} }shows \deleted{the distribution of }detected and undetected planets for an idealized survey in which all stars are at the same distance and all planets \added{are in face-on orbits and }have the same albedo. The only parameters allowed to vary here are planetary radius and semi-major axis. We see a distinct wedge-shaped pattern with sharp inner and outer working angle cutoffs (left and right, respectively), and a hard-edged contrast floor (bottom right). 

\subsubsection{Photometric condition}

For a planet to be detected, its planet/star contrast ratio must exceed the \replaced{inherent instrument floor}{coronagraph raw contrast}, $
\varepsilon >\varepsilon_{\mathrm{min}}$. We assume an optimistic LUVOIR-esque value of $\varepsilon_{\mathrm{min}} = 1\times10^{-10}$. This implicitly assumes a volume-limited survey, such that integration time per target is allotted as generously as necessary to achieve the intended signal-to-noise ratio \citep{Sta14}. 
%\added{It is implied that post-processing occurs after detection, such that planets at the raw contrast limit are robustly detected.}

\subsubsection{Obscurational condition}

Detectable planets must also have a projected separation $a_{\mathrm{proj}}$ falling outside the IWA of the coronagraph, and inside the OWA. That is, $a_{\mathrm{IWA}}<a_{\mathrm{proj}}<a_{\mathrm{OWA}}$. Both angles are set by some multiple of $\lambda/D$, where $D$ is telescope diameter. The IWA is often not actually a hard cutoff; it denotes the angular separation where the instrument sensitivity drops to 50\% its nominal value. The approximation is nonetheless reasonable\added{ \citep{Sta14}}.\footnote{\added{The sensitivity slope has a $\Delta\lambda/D$ of $\sim1$ \citep{Guy11}, which is short compared to the OWA-IWA $\Delta\lambda/D$ of 7 adopted here}.}

Projected separation is the planet's semi-major axis convolved with orbital elements\deleted{ (assuming circular orbits)},
\begin{equation}\label{eq:aproj}
a_{\mathrm{proj}} = a \sqrt{\sin^2{\xi} + \cos^2{\xi} \cos^2{i}},
\end{equation}
so $a_{\mathrm{proj}}\le a$. \added{For now we assume circular orbits, but we test this assumption below.}

The IWA limit is important for targets orbiting distant stars and/or at long wavelengths, while the OWA will pose a challenge for planets orbiting the nearest stars. \replaced{We are only concerned about the OWA limit for blind surveys, however}{The OWA is mostly of concern for blind surveys}. If we already know a planet's orbit, then we can target it at a gibbous phase with a sufficiently small projected separation (and better contrast), unless the \replaced{orbit's}{orbital} inclination is too \replaced{close to face-on}{small}. 

\subsection{Generation of planet parameters}\label{sec:rng}

%The bottom two panels of figure \ref{fig:wedge} illustrate searches of \replaced{less-unrealistic}{realistic} universes, where we draw more parameters than just $R$ and $a$ from probability density functions. This section describes these density functions. 

\subsubsection{Demographics: radius and semi-major axis}\label{radius-sma}

\citet{Pet13} showed that near 1~$R_\Earth$ and 365~days, the phase-space density of planets is approximately uniform in its natural logarithms (the actual variation was a factor of two). This distribution easily applies to semi-major axis due to Kepler's Third Law.

\added{We adopt log-uniform demographics but test the impact of this assumption below. }The normalized probability densities are
\begin{equation}\label{eq:R-pdf}
\frac{\mathrm{d}f}{\mathrm{d}(\ln{R})} = \frac{1}{\ln\left(R_{\mathrm{max}}/R_{\mathrm{min}}\right)}
\end{equation}
and
\begin{equation}\label{eq:a-pdf}
\frac{\mathrm{d}f}{\mathrm{d}(\ln{a})} = \frac{1}{\ln\left(a_{\mathrm{max}}/a_{\mathrm{min}}\right)}.
\end{equation}
These describe the likelihood of a planet having radius $R$ and semi-major axis $a$, given that the planet exists within that range of semi-major axes and radii.\added{ For a given star, the probability of a planet occurring in our playing field is about 7 in 10, as explained in section \ref{sec:numberoftargets}. Note that these ranges are broader than our adopted definition for Earth twins (see figure \ref{fig:wedge}).}

For mathematical convenience, each cell in our $3\times2$ $a$-$R$ grid (figure \ref{fig:wedge}) has a height of one $e$-folding in $R$ and a width of $e^{2/3}$ in $a$ (equal to one $e$-folding in period).\deleted{ This means we expect a constant number of planets per grid cell; similarly, it means the Earth twin occurrence rate $\eta_\Earth$ is equivalent to the Earth twin occurrence rate density $\Gamma_\Earth = \mathrm{d}\eta_\Earth/\mathrm{d}\ln R\ln P$.} The axis limits are chosen such that the cell defining Earth twins is centred at 1~AU and 1~$R_\Earth$.

\subsubsection{Orbital elements: phase and inclination}

At a given point in time, planets can be anywhere along their orbits. We assume circular orbits, so orbital phase is uniform\added{ly distributed} in $\xi \in [0, 2\pi)$, and the normalized distribution function is
\begin{equation}\label{eq:xi-pdf}
\frac{\mathrm{d}f}{\mathrm{d}\xi}  = \frac{1}{2\pi}.
\end{equation}
Meanwhile, inclination varies between 0 and $\pi/2$ and is uniform in $\cos{i} \in [0, 1]$:
\begin{equation}\label{eq:i-pdf}
\frac{\mathrm{d}f}{\mathrm{d}(\cos i)}  = 1.
\end{equation}

Inclination is an unchanging property of a planet, but orbital phase, by definition, varies as the planet orbits its star. For a given inclination and semi-major axis, there exists a maximum detectable planet/star contrast, occurring each orbit, associated with a certain orbital phase. This "optimal phase" depends on our choice of phase function model. Assuming the planet is a Lambertian reflector, the optimal phase is at the gibbous phase corresponding to $a _{\mathrm{proj}} = a_\mathrm{IWA}$, or simply the \replaced{most gibbous}{fullest} unobscured phase.

Analytically, the phase angle $\alpha$ corresponding to the optimal phase is given by substituting $a_{\mathrm{proj}} = a_{\mathrm{IWA}}$ into equation \ref{eq:aproj}, solving for $\xi$, and then substituting the result into equation \ref{eq:alpha}:
\begin{equation}\label{eq:alpha-opt}
\alpha_{\mathrm{opt}} = \sin^{-1}\left(\frac{a_{\rm IWA}}{a}\right).
\end{equation}
This equation has multiple roots; we are interested in the \deleted{solution corresponding to a waxing or waning }gibbous phase, so phase angle is $\alpha_{\mathrm{opt}} \in \left[0, \frac{\pi}{2}\right]$. 

\subsubsection{Planetary albedo}

\replaced{Albedo distributions are}{The distribution of planetary albedos is completely} unconstrained for \deleted{terrestrial }exoplanets at large separations. \replaced{To embrace}{We parameterize} this uncertainty\replaced{, we allow}{ by allowing} $A^*$ to vary over an order of magnitude, with uniform probability:

\begin{equation}\label{eq:albedo-pdf}
\frac{\mathrm{d}f}{\mathrm{d}A^*} = \frac{1}{A^*_{\mathrm{max}} - A^*_{\mathrm{min}}}.
\end{equation}

We have adopted conservative values of $A^*_{\mathrm{max}}=0.5$ and $A^*_{\mathrm{min}}=0.05$.\added{ As we will show in section \ref{sec:albedo}, the false positive rate in a blind search is insensitive to the underlying albedo distribution, or our knowledge thereof. Shrinking the albedo range decreases the false positive rate in targeted searches, but only under certain assumptions.}

\subsubsection{Distance to system}

We assume a constant density of stars out to the farthest distance probed $r_{\mathrm{max}}$, so the likelihood of a planetary system falling within a sphere of radius $r$ is proportional to $r^2$. The normalized probability density is therefore
\begin{equation}\label{eq:r-pdf}
\frac{\mathrm{d}f}{\mathrm{d}r} = \frac{3}{r^3_{\mathrm{max}}} r^2.
\end{equation}

\replaced{A net effect of introducing random planetary parameters is that u}{U}nfavourable orbits and/or greater distances shorten the time a planet spends between the inner and outer working angles. This decreases the \replaced{amount}{number} of detections, compared to a nonvarying universe (\added{cf. top and bottom panels of }figure \ref{fig:wedge}\deleted{; top panel}). The difference between figure \ref{fig:wedge}'s middle and bottom panels is due to the planet's location in its orbit, $\xi$, at the time of the image. In the middle panel, we assume that the orbit of each planet is \replaced{tracked}{known}, so we know to target stars when the planet is brightest and unobscured. On the other hand, blindly searching stars for planets is equivalent to drawing $\xi$ from its density function (eq. \ref{eq:xi-pdf}), as in the bottom panel.

\subsection{Mission parameter assumptions}

\subsubsection{Telescope}

\paragraph{Diameter}
Our notional telescope has a 10-m primary mirror, comparable to the proposed architecture\deleted{s}\added{ B} of LUVOIR and\added{ slightly greater than architecture A of} HabEx. 

\paragraph{Wavelength}
We use a wavelength of 1.0~$\micron$ to image planets in reflected starlight. This is consistent with \citet{Sta15}; they choose 1~$\micron$  as their baseline characterization wavelength due to the water vapour feature at 0.95~$\micron$.\added{ Although searching at 0.4~$\micron$ would yield more Earth twins because the IWA would be smaller, merely finding planets at this shorter wavelength is fruitless if we cannot also characterize them.}

\paragraph{Working angles}
We adopt an IWA of $3\lambda/D$ and an OWA of $10\lambda/D$, similar to the "pessimistic" case of \citet{Sta15}.

\paragraph{Contrast}
We assume the \replaced{detector}{coronagraph} has a \replaced{contrast floor}{raw contrast} of $\varepsilon_{\mathrm{min}}=1\times10^{-10}$. This threshold is often quoted as the technological goal for detection of Earth-sized planets \citep{Doo05, Rau15, Aur15}.\added{ We further assume that post-processing would provide an extra order of magnitude in contrast, enabling robust detection of planets at $\varepsilon_{\rm min}$.}

\subsubsection{Maximum survey distance}

Since we have adopted a fairly long wavelength with an accordingly large IWA, the distances at which we can probe Earth twins are limited. \deleted{Larger distances drive $a_{\mathrm{IWA}}$ outwards. }An Earth twin $r_{\mathrm{max}}$ parsecs away, orbiting at $a_{\Earth, \mathrm{max}}$, would just reach the IWA at maximum elongation; any stars beyond this point could not host \emph{detectable} Earth twins. This sets our maximum survey distance:
\begin{equation}\label{eq:rmax}
r_{\mathrm{max}} =  \frac{a_{\Earth, \mathrm{max}}}{\mathrm{IWA}} = 22.6 \;{\rm pc}.
\end{equation}

This is a \deleted{much }smaller search volume than \citet{Sta14, Sta15}, who choose \deleted{the round number of }50~pc as their maximum distance using telescope diameters of 4\textendash20~m.

\subsubsection{Number of targets}\label{sec:numberoftargets}

We assume that our survey is volume-limited, and that stars are evenly distributed across the search volume. This lets us quickly calculate the number of target stars within a sphere defined by our maximum survey distance. We use the stellar density model from \citet{Bov17}:
\begin{equation}\label{eq:bovy}
\frac{\mathrm{d}N_{*}}{\mathrm{d}V\mathrm{d}M_*} = (0.016 \; \mathrm{pc}^{-3} \;M_{\Sun}^{-1}) \;\left(\frac{M_*}M_{\Sun}\right)^{-4.7},
\end{equation}
which we integrate over [0.84, 1.15]~$M_\Sun$.

In reality, \added{not only are }$\sim$50\% of Sun-like stars\deleted{ may be} in binary pairs \citep{Bat09}\replaced{.}{, but the period distribution of binaries peaks at 10,000~days \citep{Kro01}, about the semi-major axis of Saturn.} \replaced{As we cannot block light from both stars with a coronagraph, we}{These companion stars may pose a problem for starlight suppression. Although one could improve detection yields by a factor of $\sim$2 with careful attention to coronagraph design, this is outside \replaced{our current scope}{the scope of our current paper}. We therefore} eliminate half the target\replaced{s}{ stars}; so the number of \added{target }stars in a given simulated survey is:
\begin{equation}\label{eq:N_stars}
N_{*} = \mathtt{floor}\left(\frac{1.793 \times 10^{-2}\; r_{\mathrm{max}}^3}{2}\right).
\end{equation}
This evaluates to 136 \replaced{Sun-like}{G-type} stars for $r_{\mathrm{max}}$ = 22.6~pc\replaced{: a slight overestimate}{. For comparison, \citet{Sta14} report a target list of 5449 stars within 50~pc and with spectral type A to M. Substituting these limits---excepting M-dwarfs\footnote{Only one M-dwarf, Proxima Centauri, is near enough to host Earth twins outside our adopted IWA.}---into equation \ref{eq:bovy}, we get 4937 stars}.  

\added{Simulating a realistic target list, however, is not the focus of this work. We report absolute numbers primarily as a sanity check. }\replaced{To reduce Poisson noise on certain figures and statistics in this manuscript, the number of target stars is artificially inflated.}{Indeed, most of our figures and statistics come from running 100 simulated surveys to minimize Poisson noise.} Results are otherwise unaffected by our chosen $N_{*}$.

\subsubsection{Planet occurrence rates}
To populate each star with 0 or more planets\added{ with radius $R \in [R_{\rm min}, R_{\rm max}]$ and semi-major axis $a \in [a_{\rm min}, a_{\rm max}]$}, we assume an across-the-board \deleted{Earth twin }occurrence rate density of $\Gamma = \mathrm{d}N_p/\left(\mathrm{d}\ln{R}\; \mathrm{d}\ln{P}\right) = 0.12$ planets per star per per natural logarithmic bin in period and radius \citep{Pet13, Kop18}.\added{ This corresponds to an occurrence rate, $\eta$, of about 0.7 planets per star.} In accordance with Poisson statistics, most stars have 0, 1, or 2 planets.

\begin{figure*}
   \epsscale{1.1}
    \plottwo{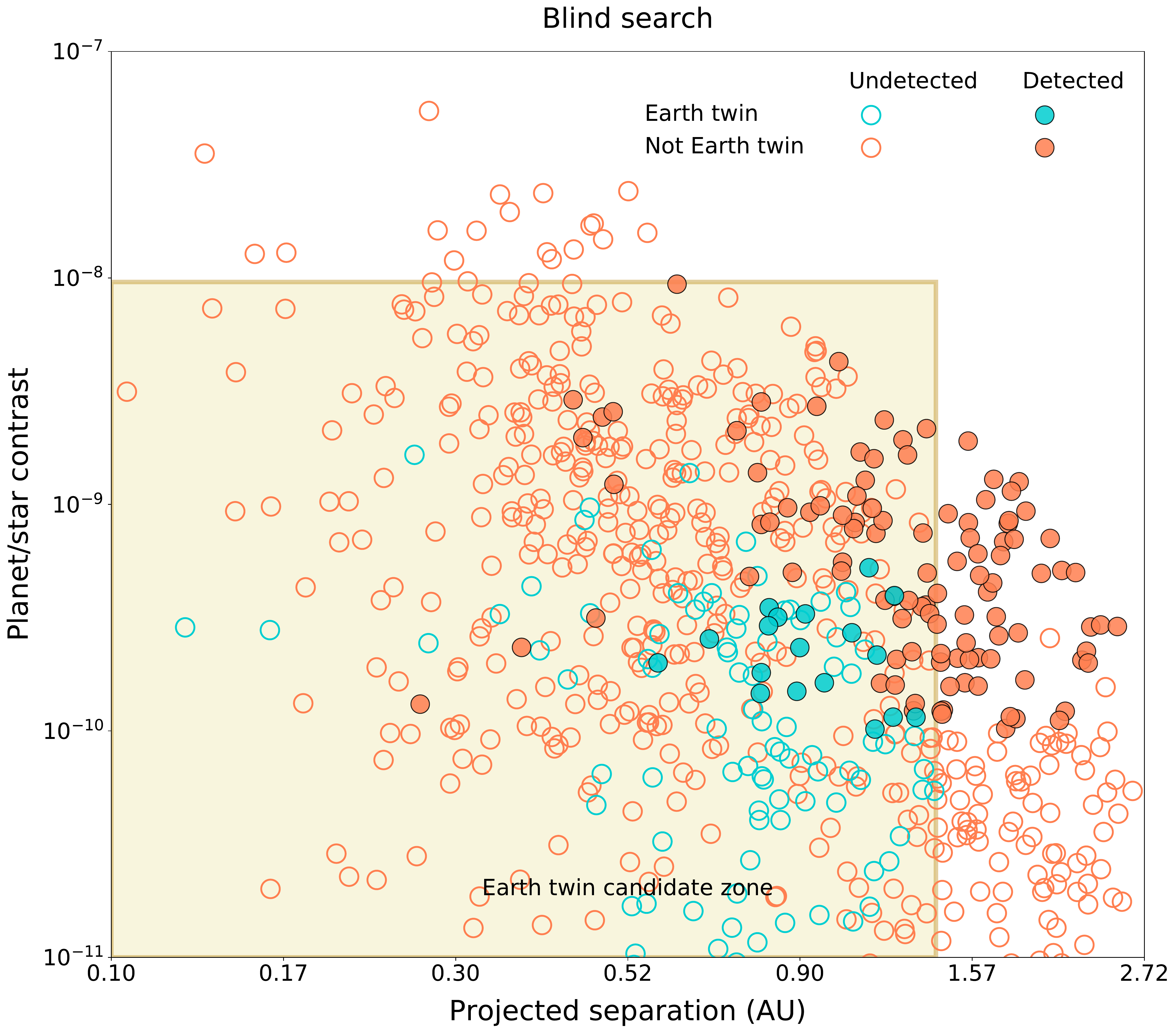}{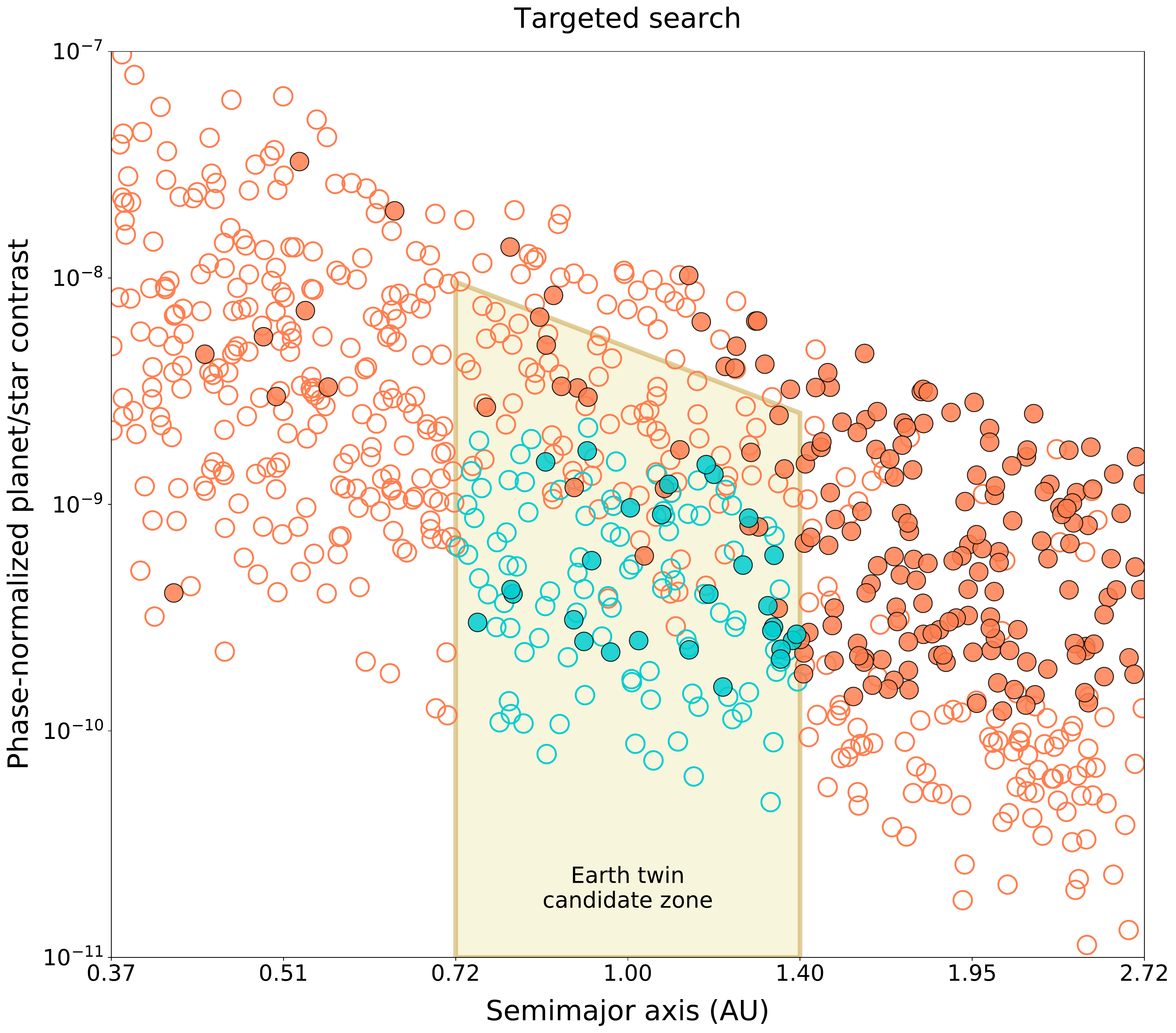}
  \caption{A comparison of survey returns in terms of the direct imaging observables, for a blind survey where planets are at random orbital phase (left), versus a survey revisiting known planets at their brightest observable phase (right). Based on a simulated survey with 10$^3$ stars; i.e., inflated by an order of magnitude, for visualization. Yellow regions show the "candidate zone" where a true Earth twin could possibly fall. Solid circles are detected planets, while empty circles are undetected planets. Blue circles represent Earth twins, and orange circles are un-Earths. Orange filled circles within the shaded region constitute planetary false positives: un-Earths masquerading as Earth twins. 
  }\label{fig:observeables}
\end{figure*}

\section{Results}
 \label{sec:earthtwindefn}
We define Earth twins\deleted{, the true positives,} in terms of planetary radius and semi-major axis. A planet orbiting a \replaced{Sun-like star}{G-dwarf} with $R \in \left[e^{-1/2}, e^{1/2}\right] R_\Earth$ and $a \in \left[e^{-1/2}, e^{1/2}\right]$ AU is an Earth twin. Note that both ranges correspond to one $e$-folding; e.g., $R_{\Earth, \mathrm{max}} = eR_{\Earth, \mathrm{min}}$. This is convenient because planetary demographics are often reported as d$N$/(dln$R$ dln$P$), so the \deleted{number of }Earth twin\replaced{s}{ rate} is simply equal to the rate density at Earth.\added{ Our Earth twins roughly encompass the "rocky" and "super-Earth" classes of \citet{Kop18}, who classify planets based on expected atmospheric chemistry. Of course, there is no evidence that all planets with the same size and orbit as Earth are anything like Earth.}

\subsection{Planetary false positive rates} \label{sec:tpr}

Locating Earth twins in figure \ref{fig:wedge} is easy---they all live in the highlighted centre grid cell on the bottom row. The problem is that \added{a single epoch of }direct imaging does not yield semi-major axis and radius, but rather, projected separation and contrast ratio. Locating Earth twins on \emph{those} axes is much trickier. We must sift through some number of un-Earthlike planets, indistinguishable from our real quarry.

We now calculate the likelihood that a planet actually is an Earth twin, given that it is detected in the contrast-separation region where an Earth twin could appear. We label this region the Earth twin candidate zone; it denotes where an Earth twin might conceivably show up in a direct imaging snapshot. The extent of the candidate zone depends on whether or not we know the planets' orbits.

\paragraph{Candidates in blind searches} \label{sec:etcz_blind}

If we know nothing about orbital phase or inclination, then the projected separation of an Earth twin on a circular orbit is at most $a_{\Earth, \mathrm{max}}$, and can be as small as 0: $0 \le a_{\mathrm{proj}} \le a_{\Earth, \mathrm{max}}$.
The maximum planet/star contrast for an Earth twin is $\varepsilon_{\mathrm{max}} = (R_{\Earth,\mathrm{max}}/a_{\Earth,\mathrm{min}})^2$; this comes from setting the apparent albedo to unity, adopting the largest Earthlike radius, and adopting the largest possible value of $\phi_L/a^2 \approx 1.44/(\pi a_{\mathrm{proj}}^2)$.\footnote{Given an observed projected separation, there is a trade-off between the semi-major axis (smaller $a$ are brighter) and orbital phase (smaller $\alpha$ are brighter). One can numerically solve for the maximum contrast ratio, which occurs at an orbital phase of about 63 degrees.}%\added{ An Earth twin with $a_{\rm proj}=0$ and/or $\varepsilon<\varepsilon_{\rm min}$ is not detectable at that epoch, but a non-detectable Earth twin is still relevant.}  

\paragraph{Candidates in targeted searches} \label{sec:etcz_phase}
If we know the \replaced{orbital phase and inclination, then $a$ can be calculated from $a_{\mathrm{proj}}$ (eq. \ref{eq:aproj}), and}{planet's orbit, then} the semi-major axis criterion for Earth twin candidacy is $a_{\Earth, \mathrm{min}} \le a \le a_{\Earth, \mathrm{max}}$.

To get the maximum contrast ratio, we divide $\varepsilon$ by its Lambertian phase function, again setting $A^*$ to unity, to compare against the stricter limit
\begin{equation}\label{eq:candidateCRb}
\varepsilon' \le\left(\frac{R_{\Earth, \mathrm{max}}}{a}\right)^2
\end{equation}
where $\varepsilon'$ is the phase-standardized contrast.

Un-Earthlike planets falling within the Earth twin candidate zone are false positives. They appear there for one or more of the following reasons:
\begin{enumerate}
\item$a < a_{\Earth, \mathrm{min}}$, but due to the planet's unknown phase and inclination, we cannot rule it out as an Earth twin in gibbous or crescent phase.
\item$a > a_{\Earth, \mathrm{max}}$, but the planet is in gibbous or crescent phase, so its projected separation \replaced{is scattered inward}{appears smaller}.
\item$R > R_{\Earth, \mathrm{max}}$, but the planet has\deleted{ sufficiently} low albedo, decreasing its planet/star contrast to something reasonable for an Earth twin.
\end{enumerate}
The degeneracy between projected separation and semi-major axis can be broken if a planet is imaged at a known orbital phase, hence ruling out \replaced{(1) and (2)}{the first two scenarios}\added{ and ameliorating the third}.

\subsubsection{Quantifying the false positive rate}
\begin{figure}
   \epsscale{1.2}
    \plotone{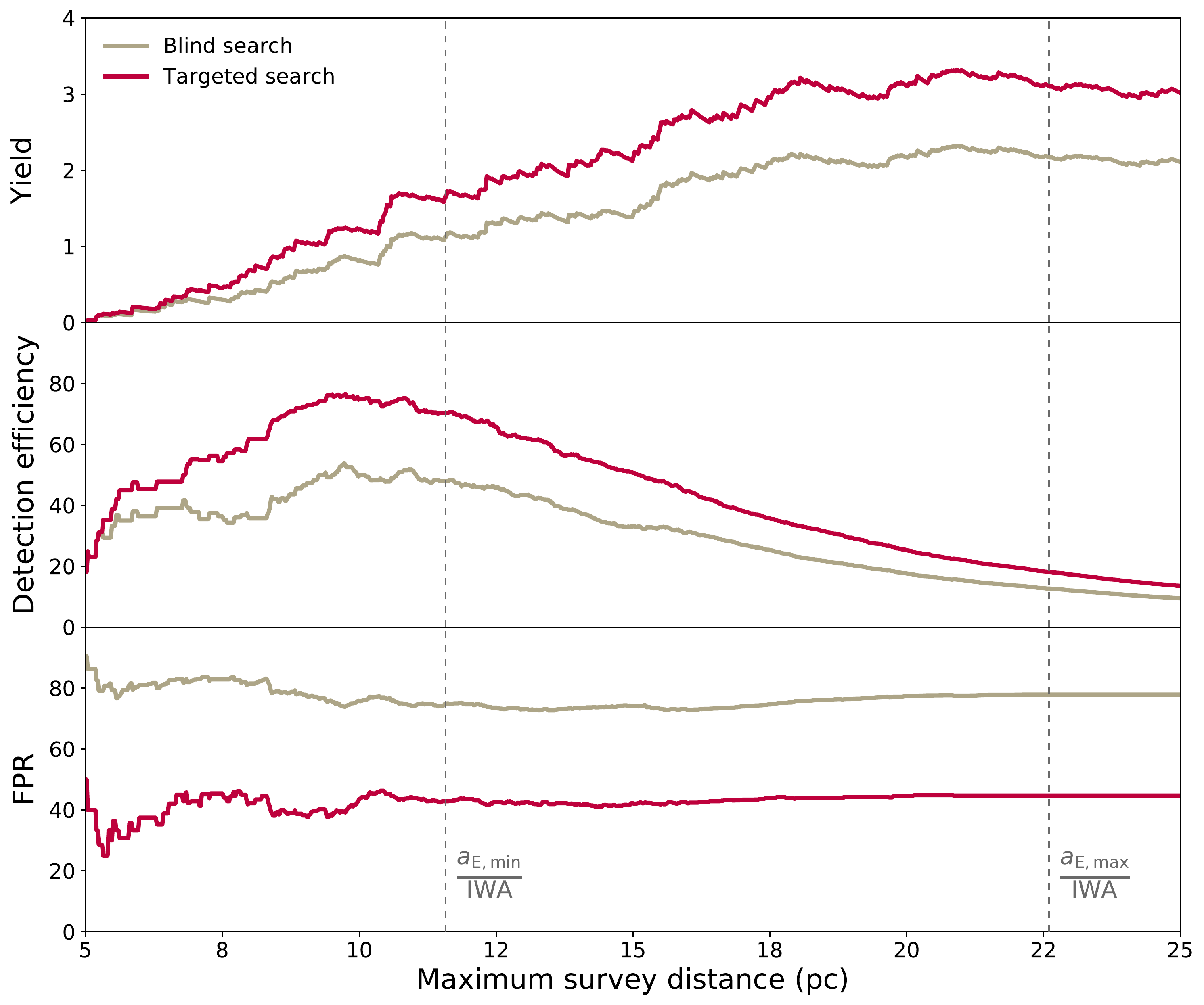}
  \caption{\added{Search volume dependence of cumulative Earth twin yield (top), cumulative Earth twin detection efficiency (middle), and cumulative false positive rate (bottom), for blind searches (grey lines) and targeted searches (red lines). Dashed vertical lines represent the distances at which a planet's projected angular separation would be just inside the IWA, if it orbited at $a_{\rm \Earth, min}$ ($r_{\rm max} = 11.6$ pc), and if it orbited at $a_{\rm \Earth, max}$ ($r_{\rm max} = 22.6$ pc). A \edit2{targeted} survey out to the leftmost dashed line would therefore \edit2{detect every Earth twin bright enough to surpass the contrast floor}, while no additional Earth twins could be detected beyond the rightmost dashed line. Calculated from 10$^5$ simulated stars, where yields are scaled to the realistic number of stars at the given $r_{\rm max}$ (equation \ref{eq:N_stars}).}
  }\label{fig:distance}
\end{figure}

We \added{essentially }count the \added{filled }dots\added{ (the detected planets)}\deleted{ (i.e.,} in figure \ref{fig:observeables}\deleted{) for detected planets} to find the false positive rate of a survey:
\begin{equation}\label{eq:tpr}
\text{FPR} = \frac{\text{[\# un-Earths]}_{\rm{det, ETCZ}}}{\text{[\# un-Earths]}_{\rm{det, ETCZ}} + \text{[\# Earths]}_{\rm det}},
\end{equation}
where the subscript ETCZ refers to a planet falling in the Earth twin candidate zone. 

Similarly, the \replaced{completeness for}{fraction of} Earth twins\added{ detected} is\added{ the number of filled teal dots to the number of teal dots:}
\begin{equation}\label{eq:completeness-earth}
f_{\text{det,} \Earth} = \frac{\text{[\# Earths]}_{\rm det}}{\text{[\# Earths]}_{\rm total}},
\end{equation}
\edit2{which we call the detection efficiency of the survey. }\added{This metric, like the false positive rate, describes the survey as a whole (cf.\ completeness from \citeauthor{Bro05} \citeyear{Bro05} being a function of a star). It is strongly dependent on the size of the search volume: visiting more distant stars becomes less efficient, despite the higher cumulative yield of planets (figure \ref{fig:distance}).}

Table \ref{tab:tpr} presents the false positive rate for blind and \replaced{known-phase}{targeted} searches. Imaging planets at their optimal orbital phases produces a lower false positive rate because the Earth twin candidate zone is smaller. Targeting planets at their optimal phases slightly improves their planet/star contrasts and minimizes the odds of missing a planet inside the IWA.

However, knowing orbits to break degeneracy is the key here, as opposed to a better-timed observation. Merely increasing the Earth twin yield via waiting for brighter and unobscured phases---without changing the candidate zone area accordingly---actually increases the false positive rate by a few percentage points to 81\%. This is because more un-Earths are also detected alongside the Earth twins.

%The false positive rate of a blind survey (\baselineFPR\%) can be improved by multiple visits. Candidates are only detectable for a fifth of their orbit on average, under our mission parameters, so subsequent visits may reveal elusive planets.

Table \ref{tab:tpr} also reports the biases in these searches. Most detected Earth twin candidates will have radii large enough such that they must have massive gaseous envelopes, making them sub-Neptunes\added{ \citep{Lop16, Ful17, Rog15}}. Phase knowledge reduces the mean radius of detected candidates from \baselineRmean~$R_\Earth$ to \baselineRmeanknown~$R_\Earth$---just outside our Earth twin box. 

The worst culprits are planets with radii too large to be Earthlike, but whose low albedos reduce their planet/star contrasts. For a blind search, we find that 67\% of Earth twin candidates will fall in this category. This statistic drops  to 47\% for targets at known phase. 

Semi-major axis degeneracy only creates false positives \replaced{if phase is unknown}{for a blind search}. In this scenario, planets orbiting exterior to $a_{\Earth, \mathrm{max}}$ make up 27\% of Earth twin candidates. Finally, planets interior to $a_{\Earth, \mathrm{min}}$ make up 9\% of candidates. Note that these categories do not add to 100\% because they are not all mutually exclusive. 

\begin{table}[h!]
\caption{False positive rates, Earth twin \replaced{completeness}{detection efficiency}, and biases in planetary radius $R$ and semi-major axis $a$ for a blind survey, versus a survey targeting planets with known orbits. False positive rate calculated via equation \ref{eq:tpr}, and detection efficiency via equation \ref{eq:completeness-earth}. Based on $10^5$ simulated stars.} \label{tab:tpr}
\begin{center}
\begin{tabular}{@{}>{\raggedright\arraybackslash}P{5.8cm} r r @{}} 
\hline
\hline
 & Blind & Targeted\\
\midrule
False positive rate (\%) & \baselineFPR & \baselineFPRknown \\
%\replaced{Earth twin completeness}{Earth twin detection efficiency} (\%) & \baselineEff &  \baselineEffknown \\
%\midrule
Mean Earth twin candidate $R$ ($R_\Earth$)& \baselineRmean & \baselineRmeanknown \\
Mean Earth twin candidate $a$ (AU)& \baselineamean & \baselineameanknown\\
 \bottomrule
\end{tabular}
\end{center}
\end{table}

\begin{table*}[htbp]
\centering
\caption{Target list sizes, number of underlying Earth twins, yields of Earth twins and un-Earths in the Earth twin candidate zone, and false positive rates under model assumptions which are relaxed one at a time. The false positive rate is quite consistent across different assumptions, despite changes in yields and target list sizes. Note that for the bottom five rows, the survey visits a dramatically different number of stars. This is due to equation \ref{eq:rmax}, where any stars with maximum Earth twin semi-major axis inside the IWA are discounted from the target list. Based on $10^5$ simulated stars, where yields are scaled to a realistic number of targets (as listed in columns 2 and 7). At this level of Poisson noise, the reported yields and false positive rates are precise to about $\pm0.2$ and $\pm2
\%$, respectively.\label{tab:sensitivity}}
{\footnotesize
\begin{tabular}{@{}l  *5{Y} c *5{Y} @{}} 
\hline
\hline
\; &  \multicolumn{5}{c}{Blind} & & \multicolumn{5}{c}{Targeted} \\
\cmidrule(lr){2-6} \cmidrule(l){8-12} 
 & \centering Stars visited \vspace{-0.1cm}& \centering $N_{\Earth}$ total & \centering $N_\Earth$ detected & \centering $N_{\rm un\Earth, ETCZ}$ detected & FPR (\%) & \;\;\;\;\;\;\;\; & \centering Stars visited \vspace{-0.1cm}& \centering $N_{\Earth}$ total & \centering $N_\Earth$ detected & \centering $N_{\rm un\Earth, ETCZ}$ detected & FPR (\%)
\tabularnewline

\midrule
 Baseline & 136 & 16.1 & \baselineYield & 7.0 & \baselineFPR & & 136 & 16.2 & \baselineYieldknown & 4.7 & \baselineFPRknown \\
FGK stars & 420 & 49.2 & 3.9 & 16.2 & 81&  & 434 & 49.5 & 10.9 & 13.9 & 56 \\
 Log-normal $R$ & 136 & 12.2 & 1.5 & 4.5 & 75&  & 136 & 12.5 & 4.2 & 2.9 & 41 \\
  Log-normal $P$ & 136 & 33.0 & 4.2 & 10.0 & 70&  & 136 & 32.9 & 11.0 & 9.4 & 46 \\
Nonzero $e$ & 136 & 16.2 & 2.1 & 7.1 & 77&  & 136 & 16.3 & 6.4 & 5.6 & 47 \\
 Log-normal $A^*$& 136 & 16.3 & 2.3 & 7.4 & 76&  & 136 & 16.3 & 6.1 & 4.3 & 41 \\
$\lambda=400$ nm & 2126 & 246.1 & 30.4 & 108.2 & 78&  & 2126 & 251.3 & 84.4 & 72.9 & 46 \\
$D=4$ m & 8 & 1.0 & 0.1 & 0.4 & 77&  & 8 & 1.0 & 0.3 & 0.3 & 46 \\
 IWA = $2\lambda/D$ & 235 & 28.1 & 3.6 & 12.1 & 77&  & 235 & 28.3 & 9.5 & 8.2 & 47 \\
  OWA = 2$\arcmin$   & 136 & 16.2 & 2.1 & 7.4 & 78&  & 136 & 16.1 & 5.4 & 4.8 & 47\\ % @ HDI FOV
 $r_{\rm max}=a_{\rm \Earth, min}$/IWA  & 18 & 2.1 & 0.9 & 2.9 & 76 & & 18 & 2.1 & 1.9 & 1.4 & 43\\
  \bottomrule
\end{tabular}}
\end{table*}

\replaced{W}{As a sanity check, w}e can estimate Earth twin yields based on a realization \replaced{with}{scaled to} a realistic number of targets, $N_*=136$\added{ G stars}. Our simulation finds $\sim$2 Earth twins in a blind search, and $\sim$5 in a \replaced{known-phase}{targeted} search. \added{Of course, our yields vary under different model assumptions, as we discuss throughout the rest of this paper.

To compare our yield results with \citet{Sta14}, we adopt their baseline mission parameters: a less forgiving telescope diameter of 8~m and IWA of 4$\lambda/D$, but a more optimistic $\lambda=550$~nm, and a larger target list of 5449 FGK stars within 50~pc. We also follow suit by fixing the occurrence rate, $\eta_\Earth$, at 0.1 planets per star across our original $a$-$R$ area. A targeted search under these assumptions finds $\sim$5 Earth twins (plus an additional $\sim$9 candidates)---consistent with the \citeauthor{Sta14} baseline yields of 4\textendash16 Earth twins for a multi-visit search (roughly equivalent to our targeted scenario), depending on astrophysical and systematic noise levels. Our Earth twin definition matters here to the extent that whereas \citeauthor{Sta14} fixed $R = 1$~$R_\Earth$, about half of our underlying Earth twins are smaller than this. Contrast ratio goes as $R^2$, so more of our simulated planets may be too faint to detect, compared to the earlier work. }

\section{Discussion}

\subsection{Model assumptions} 

We have made several simplifying assumptions throughout this numerical study. We now evaluate how damning these assumptions may be, and how they affect our results. \added{Table \ref{tab:sensitivity} summarizes our sensitivity analysis.}

\added{
\subsubsection{Search volume}

In a volume-limited survey, one must decide on a maximum survey distance, $r_{\rm max}$. There is a trade-off between detection efficiency and Earth twin yield for a given value of $r_{\rm max}$ (figure \ref{fig:distance}). Whereas our baseline survey sets the search volume such that a star at $r=r_{\rm max}$ would have all of its Earth twins obscured by the IWA, we tested a simulation where the furthest star would be complete for Earth twins. We find that this assumption reduces the targeted false positive rate from \baselineFPRknown\% to 43\%---not a very significant decrease---and this smaller volume may yield little-to-no Earth twins.}

\subsubsection{Stellar number density}

We eschew rigour for analytical convenience in estimating the number of target stars in our search volume. The stellar number density parameterization from \citet{Bov17} is not designed for lower-mass stars (\textless 1 $M_\Sun$), and will overestimate number densities in that mass region.\footnote{To get a more accurate number of G dwarfs, one should use an initial mass function below 1~$M_\Sun$ and add the result to equation \ref{eq:bovy} above 1~$M_\Sun$  (Bovy, pers. comm.). Because stellar number density is only important for absolute yield estimation, we skip this step.} Other sources of error would nevertheless dominate results.

\added{
\subsubsection{Stellar spectral type}

Our initial assumption was that all Earth twin host stars have mass $M_* = M_\Sun$. However, F, G, and K stars may be optimistically classified as "Sun-like". The semi-major axis range within which a planet would receive Earth-like insolation is farther out for F stars and closer in for K stars; we therefore expect different Earth twin detectabilities, via changes in both planet/star contrast and obscuration. Here we evaluate how a realistic distribution of stellar masses would affect our results. 

Our re-analysis is limited to stars at least as massive as the K5 spectral type. Habitable zone planets orbiting stars less massive than this---e.g., M-dwarfs---will not only have zero obliquity \citep{Hel11}, but they will also be synchronously rotating \citep{Kas93}. Their climates are likely quite alien \citep{Shi16}.

We let $M_*$ have a power law distribution, d$N$/d$M_* \propto M_*^{-4.7}$ \citep{Bov17}. We choose a normalization such that the cumulative probability equals unity in the range $M_* = [0.67, 1.6] \; M_\Sun$. Stellar luminosity is calculated by $L_*/L_\Sun = \left(M_*/M_\Sun\right)^4$. The values of $a_{\Earth, \rm{min}}$ and $a_{\Earth, \rm{max}}$ at star $q$ are then scaled by the square root of $L_{*,q}$, which effectively ignores the planetary albedo dependence on wavelength \citep{Kas93}. 

Because $a_{\Earth, \rm{max}}$ sets the edge of the the search volume (equation \ref{eq:rmax}), this means that the furthest K star probed for Earth twins is nearer than the furthest-probed F star. In other words, a star $q$ is disqualified if $a_{\Earth, {\rm{max}}, q} > r_q \times$ IWA. We find that relaxing $M_* = M_\Sun$ slightly increases the false positive rate, but this is probably not significant.
}

\begin{figure}
\epsscale{1.2}
    \plotone{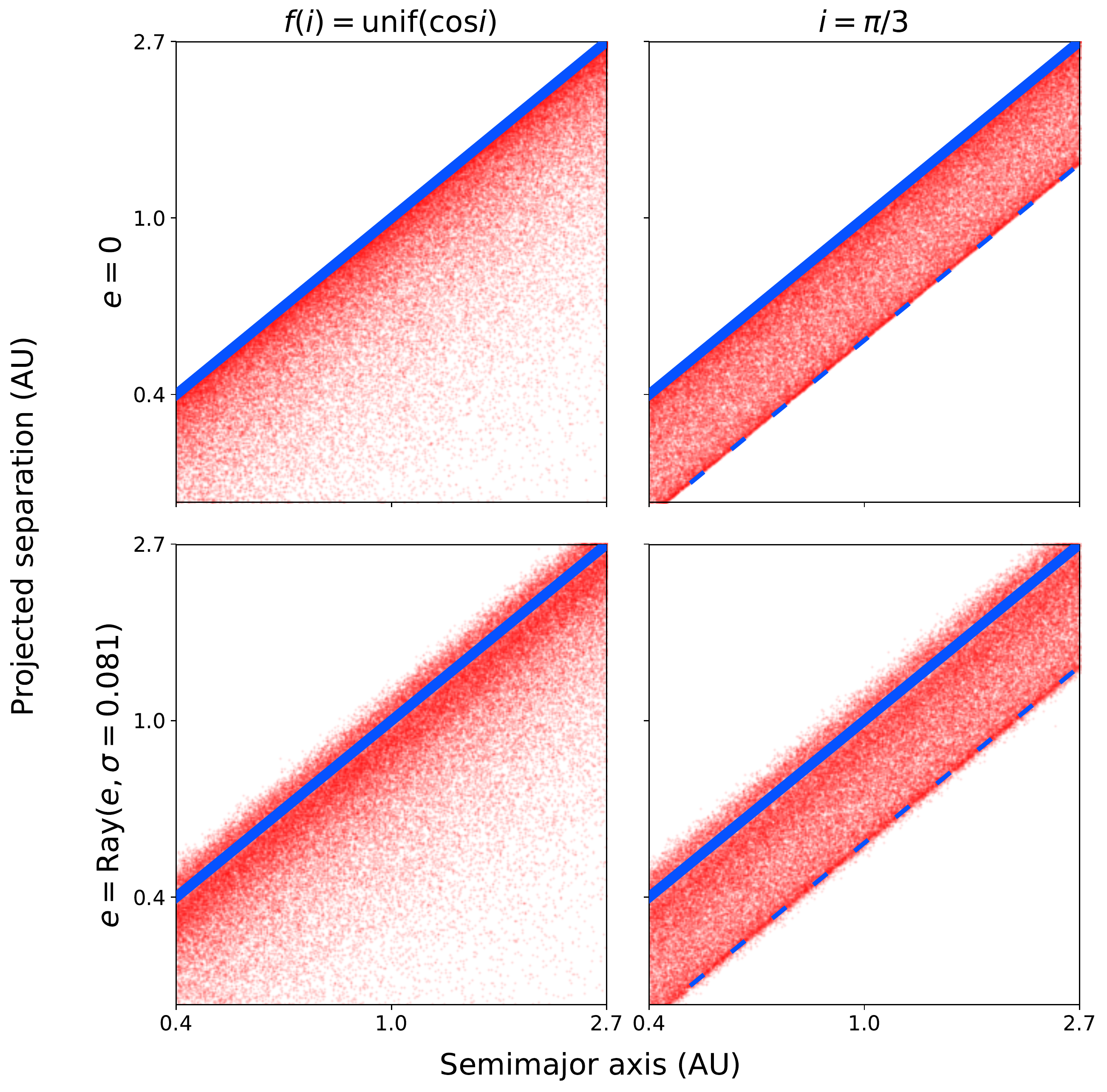}
  \caption{\added{Density scatter plots showing the distribution of projected separation with semi-major axis for different assumptions about eccentricity and inclination distributions. The solid blue line indicates 1:1 correspondence, $a_{\rm proj} = a$. The dashed blue line shows $a_{\rm proj} = a\cos i$, which is the minimum $a_{\rm proj}$ for fixed inclination and circular orbits. The distribution of $a_{\rm proj}$ with $a$ is bimodal for fixed inclination (right column) because the apparent orbital speed of the planet has minima at both $a_{\rm proj}=a$ and $a_{\rm proj}=a\cos i$, effectively piling-up planets at these four points on the orbit.}
  \label{fig:eccdist}}
\end{figure}

\subsubsection{Planetary demographics}

How appropriate is the assumption that radius and semi-major axis have log-uniform distributions? Estimating underlying distributions of exoplanets near 1~$R_\Earth$ and 1~AU is difficult because we have observed so few \replaced{real}{such} planets\deleted{ in this demographic}. Extrapolation is required, such as in \citet{Pet13}, whose flat distribution we implement in this \replaced{model}{study}. 

More recent work \citep{For14} extrapolates the distributions of radius and period using fewer assumptions than \citet{Pet13}. For planets on \textgreater100-day orbits, large radii (10~$R_\Earth$) may occur less frequently than small radii (1~$R_\Earth$), but the discrepancy is smaller than it is for shorter periods. Within the errors, however, a flat distribution does not appear to be inconsistent with \citet{For14}. 

The radius distribution of short-period planets is bimodal \citep{Ful17,Zen17}, but may be shaped by atmospheric loss via evaporation \citep{Lop16}. For planets in the habitable zone of G dwarfs in particular, the radius distribution is still poorly constrained. In any case, radius comes into the direct imaging signal as $A^*R^2$, where the apparent albedo $A^*$ is unknown. Even a bimodal distribution would likely be smeared out\added{ by albedo variance}. 

E\added{stimates of e}arth twin occurrence rate\deleted{ estimates} are \replaced{tied directly}{directly tied} to these period and radius distribution models. \citet{Pet13} present an occurrence rate $\eta_\Earth = 5.7$\%, which we divide by their Earth bin volume to get a density, $\Gamma_\Earth = 0.12$. \citet{For14} update \citeauthor{Pet13}\ to find $\Gamma_\Earth = 0.02$, smaller by an order of magnitude, while \citet{Hsu18} find $\Gamma_\Earth = 1.6$, larger by an order of magnitude. We adopt the earlier \citeauthor{Pet13}\ value because it is based on log-uniform distributions in $R$ and $a$, so we can apply a constant value of $\Gamma$ to all planets in our simulation, and still not conflict with previous work. The true occurrence rate may lie somewhere between these two results.

If $\Gamma$ is constant---that is, if planets occur at equal rates in every bin---then \replaced{completeness}{the detection efficiency} and its variation over $R$ and $a$ are divorced from the actual value of $\Gamma_\Earth$, for a volume-limited search. We are free, then, to ignore whether $\Gamma_\Earth$ is closer to 0.02 \deleted{\citep{For14} }or 0.12\deleted{ \citep{Pet13}}; its value is only needed to estimate yields. 

However, if $\Gamma$ is not constant and $\Gamma_\Earth$ is lower than its neighbouring bins\added{ \citep{For14}}, then our survey would yield more false positives. \added{Or vice versa, if $\Gamma_\Earth$ is higher than its neighbours \citep{Hsu18, Kop18}.}\deleted{Our reported false positive rates therefore represent a conservative lower estimate.}

\added{We tested how non-uniform demographics change our results by implementing log-normal distributions for both $R$ and $P$, with $\mu$ at the respective Earth value, and $\sigma$ the width of one bin. The increased abundance of Earth twins means the false positive rate is lower by a handful of percentage points, excepting the targeted scenario for a log-normal $P$ realization (since the $a$-$a_{\rm proj}$ degeneracy is trivial).}

We also prescribe an overall upper limit of 4.5~$R_\Earth$ to the planets we generate (e.g., one $e$-folding above $R_{\Earth, \mathrm{max}}$). This may miss false positives \replaced{at}{with} large\deleted{r} radii \replaced{that scatter downward in planet/star contrast, due to}{and} low \deleted{apparent }albedo, or crescent phase. Hence, again, our results give a lower limit of the false positive rate.

\subsubsection{Orbital eccentricity}

We have assumed circular orbits, but we know precious little about the eccentricity of sub-Neptunes in long-period orbits around Sun-like stars, let alone Earth twins.

\added{To test how non-zero eccentricity affects our results, we ran a simulation where eccentricity is drawn from a Rayleigh distribution with dispersion $\sigma=0.081$, as given by \citet{Sha16} for transiting planets. Although eccentricities are especially hard to measure for small planets, reports of eccentricity-period distributions consistently show peaks around $e\approx0$, for both transiting and radial velocity planets \citep{Win15}.

We find that treating $e$ as a random parameter results in a false positive rate of 77\% for a blind survey and 47\% for a targeted survey, indistinguishable from our fiducial, zero-eccentricity case. We posit that this is because inclination, not eccentricity, represents the first-order control on the distribution of projected separation with semi-major axis (figure \ref{fig:eccdist}). We therefore conclude that our analysis is robust to the assumption of circular orbits.}

\begin{figure}
\epsscale{1.17}
    \plotone{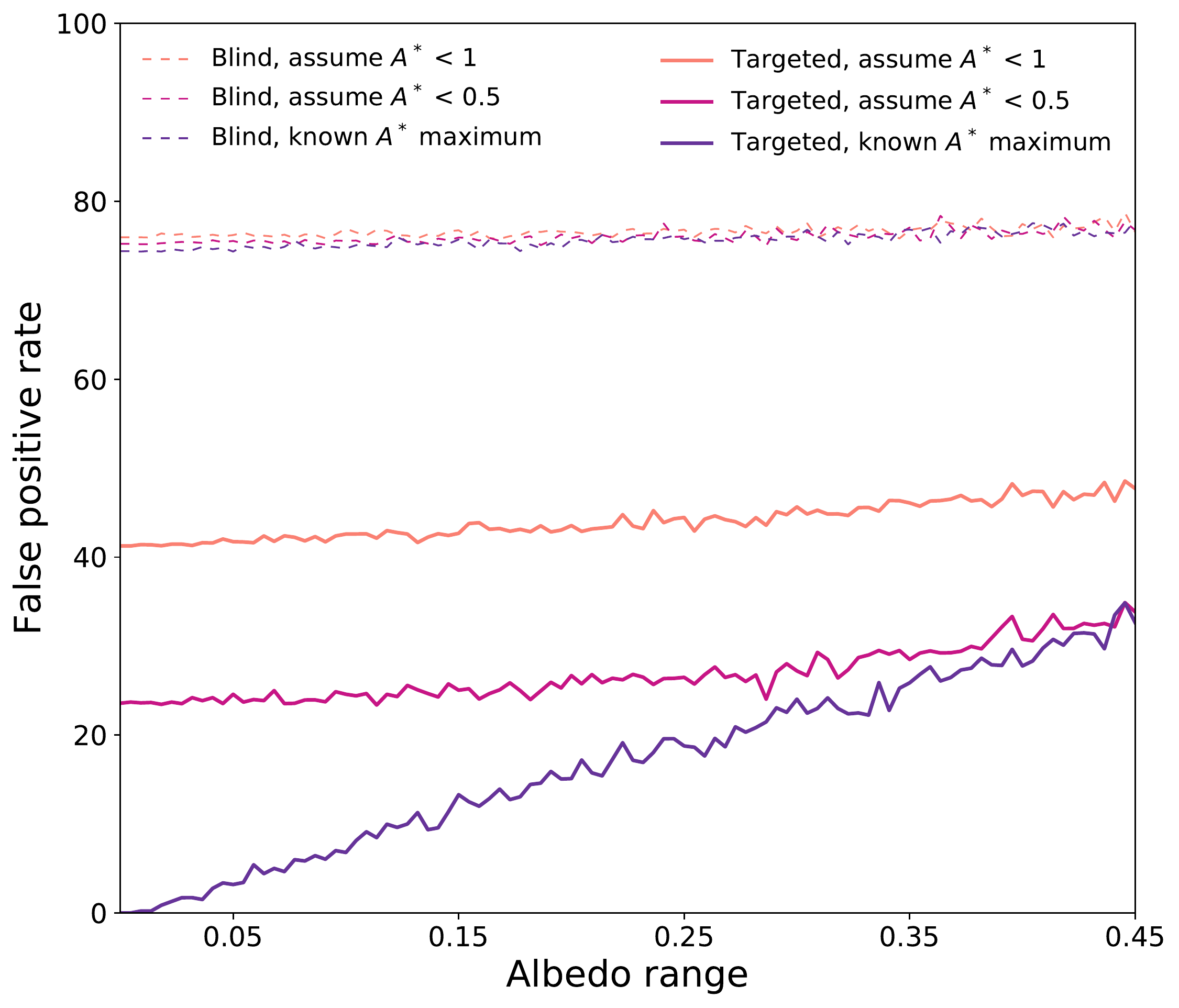}
  \caption{Effect of albedo distribution---and our knowledge thereof---on the false positive rate, or, the odds of an Earth twin candidate being an un-Earth. The $x$-axis is the range within which albedo is allowed to randomly vary \emph{in the model}: the greatest range corresponds to $A^* \in [0.05, 0.5]$, and the smallest range to $A^* = 0.3$. Dashed lines represent searches for planets at random phase, while solid lines represent \added{targeted }searches\deleted{ with prior phase knowledge}. Colours show different assumed maximum albedos (e.g., the value of $A^*$ in equation \ref{eq:candidateCRb}). Noise in this figure is due to model Poisson noise: because $A^*$ is \deleted{re}generated\added{ anew} for each planet per albedo range increment, sometimes planets will be assigned new $A^*$ values sufficiently low to diminish their planet/star contrasts below the instrument floor, which renders them undetectable. The false positive rate in a blind search is insensitive to the underlying albedo distribution, or our knowledge thereof (dashed lines). \replaced{Searching for planets at known orbital phase incurs false positive rates that are much lower, but still approximately 50\%}{Targeted searches still have false positive rates of at least 1 in 2}, unless all planets have the same albedo (albedo range of 0) and we know that universal albedo \textit{a priori} (solid purple line).\label{fig:albedo}}
\end{figure}

\begin{figure}\epsscale{1.17}
    \plotone{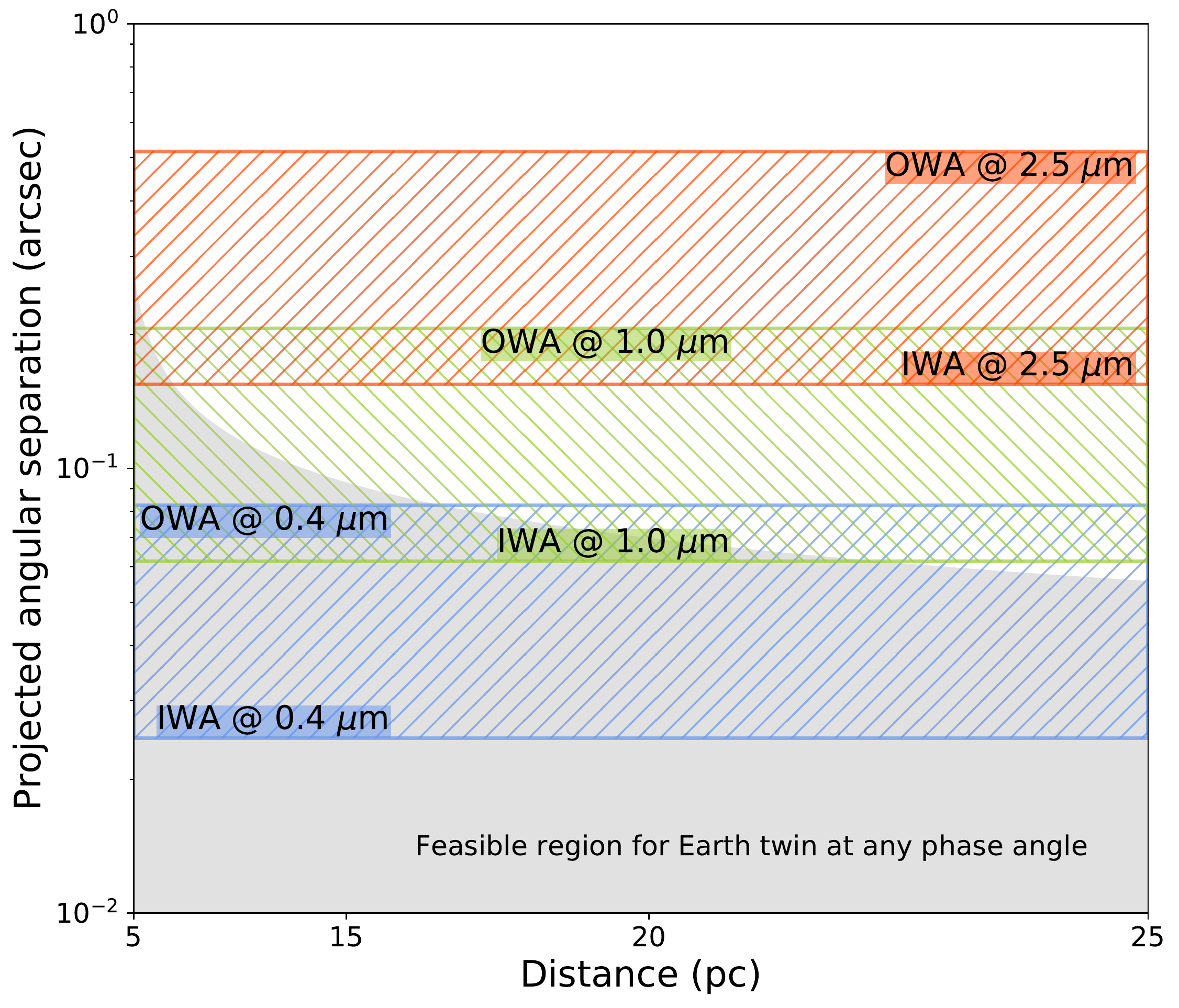}
  \caption{Inner and outer working angles at various wavelengths (horizontal lines), for a 10-m telescope with an IWA of 3$\lambda/D$ and an OWA of 10$\lambda/D$. Hatched regions represent where an exoplanet would be unobscured. The angular projected separation for an Earth twin, as a function of distance, is shown by the grey region. Targets are only visible at some wavelength if the grey swath intersects a wavelength's working angle box. The hatched regions have little-to-no overlap, meaning that no planets can be simultaneously imaged from 0.4 to 2.5~$\micron$, and a full spectrum can only be stitched together for the very nearest and most inclined planets. Because the $x$-axis is scaled to constant volume per centimetre, this demonstrates that the vast majority of Earth twins have too tight a projected angular separation for longwave characterization.\label{fig:workingangles}}
\end{figure}

\subsubsection{Phase function and albedo}\label{sec:albedo}

We have adopted the Lambertian phase curve throughout our analysis. \deleted{Under this assumption, the light reflected by the planet's atmosphere is diffuse---it scatters in all directions. }In reality, however, a planet's phase function will differ from the Lambertian model \citep{Bur09,May16}---for example, Titan is strongly forward-scattering and appears brighter at larger phase angles \citep{Gar17}. Our ignorance of exoplanet phase functions is largely encapsulated in the apparent albedo, which we allow to vary by an order of magnitude.

As we have stressed throughout this work, the albedo distribution of rocky planets is wholly unconstrained. Moreover, $A^*$ may change as we observe different regions of the planet \citep{Cow17}. %This variation is controlled by (i) the planet's rotation about its axis, (ii) the obliquity of that axis, and/or (iii) weather and seasons \citep[see][]{Cow17}.

As for our assumptions about $A^*_{\mathrm{min}}$ and $A^*_{\mathrm{max}}$, hot Jupiters exhibit more than an order of magnitude range in albedo \citep{Hen17}, despite being relatively simple planets: similar mass, size, composition, etc. There is therefore reason to believe that smaller, cooler planets, which are inherently more diverse, will exhibit a variety of different albedos.

In figure \ref{fig:albedo}, we present the Earth twin false positive rate as a function of the underlying range of apparent albedo. The problem of unknown albedo is twofold: not only do we not know the albedos of individual planets, but we do not even know the albedo \emph{distribution} of planets at 1 AU. Therefore, we are left with (i) our best guess for $A^*_{\mathrm{max}}$ (which affects the \added{extent of the }Earth twin candidate zone\deleted{ extent}), as well as (ii) our luck in nature's range of $A^*$ being on the small side. 

The Earth twin false positive rate varies with both of these estimates. If every candidate planet had the same albedo and phase function \emph{and} we knew the universal albedo and phase function \textit{a priori}, then---and only then---would a \replaced{phase-informed}{targeted} search return a 0\% false positive rate, since the radius-albedo degeneracy would be broken. As the universe's underlying distribution widens, however, our knowledge of the albedo maximum gives us less and less of an advantage.

\added{Table \ref{tab:sensitivity} shows that adopting an underlying normal distribution for $A^*$ ($\mu=0.3$, $\sigma=0.1$) also reduces the false positive rate of a targeted search, in a similar way to shortening the range of $A^*$.}

\begin{figure*}
    \epsscale{1.17}
    \plotone{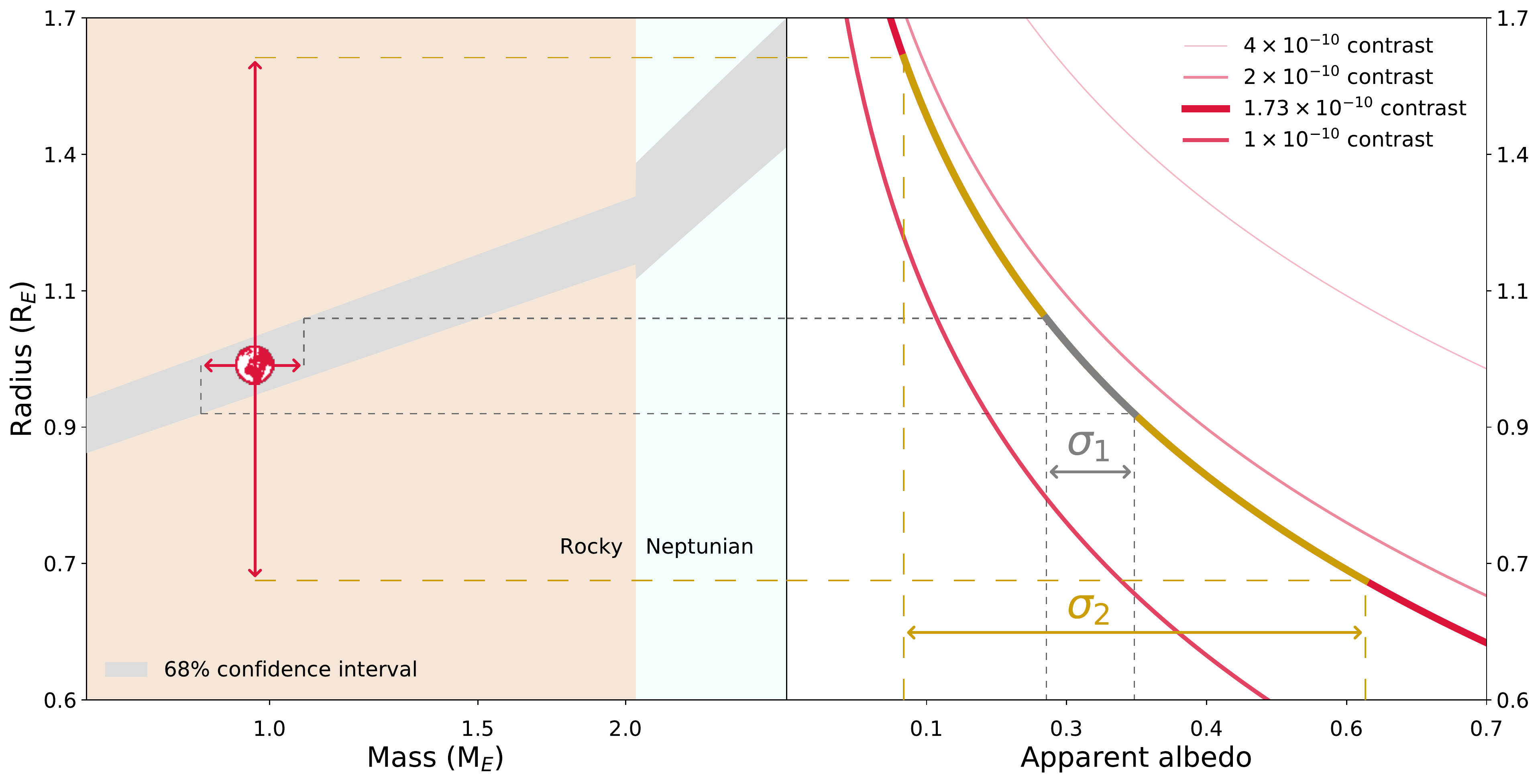}
  \caption{\textit{Left:} Planetary mass-radius relation from \citet{Che16}. Grey swath shows 68\% confidence interval. Horizontal error bars are the hypothetical mass measurement error, here set at a very optimistic value of 10\% (e.g., using \replaced{astrometry}{1 cm/s precision radial velocity; \citeauthor{Pla18} \citeyear{Pla18}}). \edit2{Vertical error bars show a hypothetical radius constraint retrieved from a Rayleigh scattering spectrum \citep{Fen18}. }The dashed lines show $1\sigma$ radius constraints\edit2{, where grey lines correspond to the mass constraint and ochre lines correspond to the spectral retrieval constraint}. \textit{Right:} radius-albedo degeneracy at a constant planet/star contrast of 1.73$\times10^{-10}$ (bold line), which corresponds to an Earth twin at quadrature and 1~AU separation. Other lines show different planet/star contrasts for the same phase and separation. The error on radius, as estimated from mass\edit2{ or from spectral retrieval}, directly propagates to an error on albedo. We might then estimate albedo to within roughly $\pm0.05$\edit2{ ($\sigma_1$) or $\pm0.25$ ($\sigma_2$), respectively,} for planets with Earthlike albedo and radius.\label{fig:mass}}
\end{figure*}

\subsubsection{Wavelength and working angles}\label{sec:OWA}

\replaced{We have mentioned, but not yet stressed, that i}{I}nner and outer working angles depend directly on imaging wavelength. Shorter wavelengths will tighten the working angles, while longer wavelengths will push them to wider separations. The wavelength we choose to work with thus affects which planets are obscured and which are not. Our \replaced{assumption}{adoption} of 1.0~$\micron$ \replaced{entails}{dictates} that planets are obscured more often than the 0.55-$\micron$ assumption of \citet{Sta14}. Indeed, \citet{Sta15, Sta16} require that planets are simultaneously detectable at 0.55 and 1.0~$\micron$.\added{ Regardless, the false positive rate is roughly insensitive to both the wavelength and the working angles themselves (table \ref{tab:sensitivity}).}

If we want to spectroscopically characterize the atmospheres of planets we detect\deleted{ (i.e., do useful science)}, then we require observations at multiple bands. For full characterization, we would hope for a spectrum ranging from 400~nm in the shortwave (Rayleigh scattering), to 2.5~$\micron$ in the longwave (greenhouse gas absorption, e.g. methane).

Directly imaging a planet at multiple wavelengths is not trivial, however, due to chromatic working angles. We illustrate this in figure \ref{fig:workingangles} by showing the projected angular separations at which an Earth twin might appear, overlain by the working angles at some different wavelengths.

If we want to \textit{simultaneously} detect a planet at multiple wavelengths, then the regions bounded by the relevant IWAs and OWAs \emph{and} the planet's angular separation must all overlap somewhere. As figure \ref{fig:workingangles} shows, this is unfortunately not achievable for 400~nm and 2.5~$\micron$, \replaced{using an OWA of $10\lambda/D$ and an IWA of $3\lambda/D$}{if OWA $=10\lambda/D$ and IWA $=3\lambda/D$}.\added{ Parallel coronagraphs, with different IWAs and OWAs, are a possibility for imaging more planets at such a range of wavelength bands.}

Thus we may be forced to attempt stitching together observations taken at different phases\added{, at least for planets on inclined orbits}. This raises practical challenges, since $\phi(\alpha)$ varies with wavelength; phase variations are likely chromatic \citep{Cah10,May16}. Further---and this extends to all of Earth twin spectroscopy---we are chasing moving targets. The integration time required to characterize an Earth twin could be on the scale of months \citep{Rob16}, and a planet on a 1-AU orbit will surely move during this time.\footnote{\added{A planet with $a=1$~AU at $r=10$~pc would move 5 pixels over a 30-day integration, assuming a Nyquist-sampled pixel scale and a 10-m telescope. The same planet at $r=20$~pc would move 2.5 pixels.}} \replaced{\textit{A priori} orbital knowledge would therefore be prerequisite, as opposed to just useful}{It may therefore be necessary to acquire orbital constraints before obtaining spectra}. 

Regardless, even with snapshots at several orbital phases, figure \ref{fig:workingangles} illustrates that only the nearest ($r \lesssim 9$ pc) Earth twins are possibly observable both at 400~nm and 2.5~$\micron$. Of the simulated Earth twins detectable at 400~nm, 23.9\% are detectable at 1.0~$\micron$ at any phase, and only 0.6\% at 2.5~$\micron$. 

One debatable solution is to use a starshade, rather than a coronagraph, to obtain spectra of Earth twin atmospheres. The IWA of a starshade depends on the starshade radius divided by the starshade-telescope distance, and its OWA is simply the field of view. This results in a greater unobscured range of separations. Starshades also have greater bandwidth, so obtaining a full spectrum requires fewer passes. However, because starshade slew time is long, fewer stars can be targeted\added{, and starshades themselves pose different technical challenges}. 

\subsubsection{Multiple observations}

\edit2{Our blind search model assumes one observation per star, while our targeted search assumes either precursor orbit constraints, or enough direct imaging visits per star to fully constrain planetary orbits. A realistic mission will fall between these endmembers---at a given point, we may have visited a star more than once, yet possibly not enough times to precisely know the semi-major axis of the hosted planet(s). This raises an interesting question: how does the false positive rate change with each additional visit to the same star? The answer requires knowing the most efficient timing of visits, an important area of future research. For now, we posit that our false positive rates reported for the blind and targeted scenarios represent upper and lower bounds, respectively.}

\subsection{Breaking the radius-albedo degeneracy}

\edit2{We consider two possible routes to constraining planetary albedo (figure \ref{fig:mass}). }One \deleted{ possible} route \deleted{to constraining planetary albedo }is to choose targets whose masses are known from radial velocity or astrometry surveys \citep{Pla18, Sha18, Ben15, Fis16, Wei16}. We can use a mass-radius relation \citep[e.g.,][]{Che16} to estimate the planet's radius from its mass. This is a risky endeavour, as current mass-radius relations are necessarily for short-period planets and therefore may not be representative of Earth twins. A corollary benefit of targeting known-mass planets is that their orbits would have been constrained along with mass. This would inform us of which stars to target and when to look. 

\edit2{The second route takes advantage of Rayleigh scattering. \citet{Fen18} showed that modeled Rayleigh scattering spectra are independent of surface albedo, and could therefore constrain radius. In theory, if we measure the Rayleigh scattering spectrum of a planet at known phase, then we can estimate its radius.}

\edit2{This retrieval is more complicated for an atmosphere with clouds. However, the longer atmospheric path-lengths at crescent phase mean that surface and cloud scattering are less important at these phase angles. Thus, reflected light at crescent phase is---in principle---closer to pure Rayleigh scattering, and hence might constrain radius, even for cloudy atmospheres.}

Figure \ref{fig:mass} shows that a 10\% constraint on mass would propagate to approximately a $\pm 0.1 \; R_\Earth$ constraint on radius and a $\pm0.05$ constraint on albedo, for Earthlike planets at 1~AU\edit2{, and that a 50\% radius constraint from a Rayleigh scattering spectrum would propagate to an $\pm0.25$ constraint on albedo}. A precise value of $\sigma_{A^*}$ is not reported because this error would be dominated by systematic errors; e.g., using a mass-radius relationship for short period planets.

\section{Conclusions}

\begin{table*}[t]
\caption{Definitions of symbols used in this text.\added{ Listed values correspond to the fiducial case; many of these parameters are varied in our sensitivity analysis.}} \label{tab:params}
\begin{center}
% symbol value min max units description equation
\begin{tabular}{@{} l r r l >{\raggedright\arraybackslash}P{7cm} >{\raggedright\arraybackslash}P{2.7cm} r@{}} 
\hline
\hline
\multicolumn{7}{c}{\textbf{Randomly-generated planetary parameters}}\\
Symbol &  Min & Max & Units & Description & Probability distribution & Eqn.\\
\midrule
$R$ &  0.6 & 4.5 & $R_\Earth$ & Planetary radius & Uniform in ln($R$)& \ref{eq:R-pdf}\\ 
$a$ &  0.37 & 2.72 & AU & semi-major axis & Uniform in ln($a$) & \ref{eq:a-pdf}\\ 
$\xi$ &  0 & $2\pi$ & rad & Orbital phase & Uniform&\ref{eq:xi-pdf} \\
$i$ & 0 & $\pi/2$& rad & Orbital inclination & Uniform in cos($i$)&\ref{eq:i-pdf} \\
$A^*$ & 0.05 & 0.5 & \multicolumn{1}{c}{-} & Planetary apparent albedo & Uniform&\ref{eq:albedo-pdf} \\
$r$ & 0 & 22.6 & pc & Distance between star system and observer & Uniform in $r^2$&\ref{eq:r-pdf} \\ 

\hline
\hline
\multicolumn{7}{c}{\textbf{Derived planetary parameters}} \\
Symbol &  &  & Units & Description & & Eqn.\\
\midrule

$\varepsilon$ &  & &  \multicolumn{1}{c}{-} & Planet/star contrast ratio & &\ref{eq:CR} \\
$\phi_L(\alpha)$&   &  & \multicolumn{1}{c}{-} & Lambertian phase function& & \ref{eq:phi} \\
$\alpha$ &  &  &rad & Phase angle between planet and observer && \ref{eq:alpha} \\
$a_{\mathrm{proj}}$ &  &   & AU & Projected separation && \ref{eq:aproj} \\
$\varepsilon'$ &  & &  \multicolumn{1}{c}{-} & \multicolumn{2}{l}{Phase-normalized planet/star contrast ratio} &\ref{eq:candidateCRb}\\

\hline
\hline
\multicolumn{7}{c}{\textbf{Free model parameters}} \\
Symbol &  Value &  & Units & Description &  & \\
\midrule

$\varepsilon_{\mathrm{min}}$ & $1.0\times10^{-10}$ &   & \multicolumn{1}{c}{-} & Coronagraph raw planet/star contrast ratio \\
$R_{\Earth, \mathrm{min}}$ & 0.6 &  & $R_{\Earth}$ & Minimum radius for Earth twin classification \\
$R_{\Earth, \mathrm{max}}$ & 1.6 &  & $R_{\Earth}$ & Maximum radius for Earth twin classification \\
$a_{\Earth, \mathrm{min}}$ & 0.72 &  & AU & \multicolumn{3}{l}{Minimum semi-major axis for Earth twin classification} \\
$a_{\Earth, \mathrm{max}}$ & 1.40 &  &  AU & \multicolumn{3}{l}{Maximum semi-major axis for Earth twin classification} \\
$\Gamma_\Earth$ & 0.119 &  &  nat$^{-2}$ & Earth twin occurrence rate density \\
$M_*$ & 1 &  & $M_{\Sun}$ & Mass of host star\\
$D$ & 10 &  & m & Telescope primary aperture diameter\\
$\lambda$ & 1.0   & & $\micron$ & Imaging wavelength\\
$N_{\mathrm{in}}$ & 3   & & \multicolumn{1}{c}{-} & \multicolumn{3}{l}{Number of $\lambda/D$ at coronagraph inner working angle}\\
$N_{\mathrm{out}}$ & 10  &  & \multicolumn{1}{c}{-} & \multicolumn{3}{l}{Number of $\lambda/D$ at coronagraph outer working angle} \\
 \bottomrule
\end{tabular}
\end{center}
\end{table*}

We have performed Monte Carlo simulations of reflected light direct imaging surveys adopting a simple telescope model. Our main finding is: if we image stars at random, $\sim$\baselineFPR\% of the detected planets that appear Earthlike in separation and planet/star contrast will in fact not be Earth twins. Meanwhile, $\sim$88\% of Earth twins go undetected within our search volume of 22.6~pc\edit2{, although this depends on model assumptions; namely, the maximum survey distance in our volume-limited survey}. 

We can double the chances that detected Earth candidates are true Earth twins---and triple the chances of seeing Earth twin planets, on average---by \replaced{patiently waiting for}{only targeting known planets}. Yet even then we cannot do better than a $\gtrsim$50\% false positive rate, as our capacity to know whether a planet is an Earth twin is set by our knowledge of the albedo distribution of rocky planets at large semi-major axes.\edit2{ These two estimates of the false positive rate represent endmember search scenarios, in which we either know nothing or everything about the orbits of the imaged planets. The false positive rate of a realistic direct imaging mission would fall in between these values.}

\added{Our results are robust to working angle geometry (including imaging wavelength), to the assumption of non-circular orbits, to the inclusion of F and K stars, and to the underlying radius, period, and albedo distributions of planets.}

Breaking the radius-albedo degeneracy should be a focus of research before choosing Earth twin candidates for costly spectroscopic characterization. We may be able to constrain a planet's radius from its mass, motivating cooperation between direct imaging and radial velocity and astrometry.

\acknowledgments
This work is supported by the McGill Space Institute and the Technologies for Exo-Planetary Science training program. We show our gratitude to the LUVOIR STDT for their useful conversation, to the Exoclipse meeting attendees and organizers for the feedback received on a preliminary presentation of this work, to M. Marley and T. Robinson for enlightening discussion on the radius-albedo degeneracy, and to an anonymous referee, whose comments greatly improved the scientific quality of this manuscript. CMG thanks T. Bell, L. Dang, and D. Keating for additional discussion. Lastly, NBC acknowledges the support of the Canadian Space Agency towards attending LUVOIR meetings.
%\newpage
\bibliography{earthtwins}

 \listofchanges
\end{document}